\documentclass[preprint,prl,superscriptaddress]{revtex4}

\usepackage{dcolumn}
\usepackage{amsmath}

\usepackage{graphicx}

\def\eref#1{(\ref{#1})}
\def\d{{\rm d}}

\def\grad{\nabla}
\def\vr{{\bf r}}

\newlength{\figurewidth}
\begin{document}

\title{Optimal design, robustness, and risk aversion}
\author{M. E. J. Newman}
\affiliation{Santa Fe Institute, 1399 Hyde Park Road, Santa Fe, NM 87501}
\author{Michelle Girvan}
\affiliation{Santa Fe Institute, 1399 Hyde Park Road, Santa Fe, NM 87501}
\affiliation{Department of Physics, Cornell University, Clark Hall,
Ithaca, NY 14853--2501}
\author{J. Doyne Farmer}
\affiliation{Santa Fe Institute, 1399 Hyde Park Road, Santa Fe, NM 87501}

\date{February 19, 2002}

\begin{abstract}
  Highly optimized tolerance is a model of optimization in engineered
  systems, which gives rise to power-law distributions of failure events in
  such systems.  The archetypal example is the highly optimized forest fire
  model.  Here we give an analytic solution for this model which explains
  the origin of the power laws.  We also generalize the model to
  incorporate risk aversion, which results in truncation of the tails of
  the power law so that the probability of disastrously large events is
  dramatically lowered, giving the system more robustness.
\end{abstract}


\maketitle

In a series of recent papers, Carlson and Doyle~\cite{CD99,CD00,DC00} have
proposed a model for designed systems which they call ``highly optimized
tolerance'' or HOT.  The fundamental idea behind HOT is that systems
designed for high performance naturally organize into highly structured,
statistically unlikely states that are robust to perturbations they were
designed to handle, yet fragile to rare perturbations and design flaws.  As
an example they consider an idealized model of forest fires~\cite{CD00}.
In this model a forester is charged with finding the optimal distribution
of the trees on a grid so as to maximize tree harvest in the face of
occasional fires that burn complete connected clusters of trees and are
started by sparks that arrive with a given spatial distribution.  They find
that optimizing the harvest, or yield, for the model gives rise to a
segmented forest consisting of contiguous patches of trees separated by
firebreaks, and that the resulting distribution of fire sizes usually
follows a power law.  While this type of configuration does typically
achieve very good yields, the system is also fragile in the sense that
perturbations to the firebreaks or changes in the spark distribution can
lead to substantially sub-optimal performance.  They argue that these are
pervasive phenomena: high-performance engineering leads to systems that are
robust to stresses for which they were designed but fragile to errors or
unforeseen events.

In this paper we argue that simple yield maximization is problematic even
if there are no errors in firebreaks or changes in the spark distribution.
Because the power-law distributions generated by yield maximization have
fat tails, disastrously large forest fires occur with non-negligible
frequency---far greater frequency than one would expect from intuition
based on normal distributions.  This idea, that yield optimization can lead
to ruinous outcomes, is not new.  For the classic problem of gambler's
ruin, for example, it is well known that optimizing total return leads to
ruin with probability one.  By contrast, if one is willing to accept
suboptimal returns it is possible to construct gambling strategies that are
immune to ruin~\cite{CT91}.  Applying similar ideas in the present context,
we show that a risk-averse engineer who is willing to accept some loss in
average system performance can effectively limit the large deviations in
the event size distribution so that disasters are rare.  We call this
variation on the HOT theme ``constrained optimization with limited
deviations'', or COLD.  By avoiding total ruin, a COLD design is more
robust than a HOT one, even in a world of perfect error-free optimization.

To demonstrate the difference between HOT and COLD we first revisit the HOT
forest fire model.  We give an analytic solution for the model which shows
that the distribution of fire sizes does indeed follow a power law, the
cumulative distribution having a exponent $-(1+1/d)$, where $d$ is the
dimensionality of the system.  Using a fast percolation
algorithm~\cite{NZ00} we perform numerical simulations that confirm the
value of this exponent.  We then generalize our solution to include risk
aversion in the design, thereby breaking the power law scaling and
dramatically reducing the frequency of disastrously large events.

Following Refs.~\onlinecite{CD99} and~\onlinecite{CD00} then, we consider a
forest divided into a large number of regions or patches, with firebreaks
between them that prevent the spread of fire from one patch to another.
Although the original forest fire model was based on a lattice, the model
we consider is a continuum one, since this makes the mathematical treatment
more tractable.  For large system sizes, we expect the behavior of this
continuum model to converge to that of the lattice model.

It is assumed that during the lifetime of the forest a single spark lands
at a random position $\vr$ and starts a fire that burns the surrounding
patch.  Let us denote the area of this patch by~$s(\vr)$.  The forest is
then harvested, giving a yield equal to the area of the remaining forest.
In units where the total area of the forest is one, the yield is $1 -
s(\vr) - F$, where $F$ is the cost in terms of yield of constructing the
firebreaks.

Because dimensionality is an important property of HOT systems, we consider
the model for general dimension~$d$.  If the cost of constructing
firebreaks is~$a$ per unit length (or per unit surface area for $d>2$),
then the cost of the firebreak surrounding a patch~$m$ is
$agds_m^{(d-1)/d}$, where $s_m$ is the value of $s(\vr)$ in patch~$m$ and
$g$~is a geometric factor of order~1 that depends on the geometry of the
lattice and the shape of the patches.  In the lattice version of the forest
fire model, $a$~is simply equal to the lattice parameter (i.e.,~the
nearest-neighbor spacing), but in the continuum model we are at liberty to
give $a$ any value we feel to be appropriate.  As we will see, as long as
$a$ is finite its value does not affect the shape of the distribution of
fire sizes.

Because $s(\vr)$ is constant inside each patch, the integral of $1/s(\vr)$
over any patch is identically~1, and hence, summing over all patches, the
total area occupied by firebreaks is
\begin{equation}
F = agd \sum_m s_m^{(d-1)/d} 
  = agd \sum_m s_m^{(d-1)/d} \int_m {\d^d r\over s(\vr)}
  = agd \int s(\vr)^{-1/d}\>\d^d r.
\label{perimeter}
\end{equation}
Letting the normalized probability distribution of sparks be
$p(\vr)$, the mean yield is then
\begin{equation}
Y = 1 - \int p(\vr) s(\vr) \>\d^d r - agd \int s(\vr)^{-1/d} \>\d^d r,
\label{yield}
\end{equation}
where the integrals run over the entire area of the forest.

To find the maximum yield with respect to the patch sizes $s(\vr)$, we set
the functional derivative $\delta Y/\delta s(\vr) = 0$,
giving
\begin{equation}
s(\vr) = \biggl[ {ag\over p(\vr)} \biggr]^{d/(d+1)},
\label{sr}
\end{equation}
for all~$\vr$.  The optimal yield is then given by substituting back into
Eq.~\eref{yield} to get
\begin{equation}
Y_{\rm opt} = 1 - (d+1) (ag)^{d/(d+1)} \int p(\vr)^{1/(d+1)} \>\d^d r,
\end{equation}
and the optimal number of patches is
\begin{equation}
n = \int {\d^d r\over s(\vr)} = (ag)^{-d/(d+1)} \int p(\vr)^{d/(d+1)}
    \>\d^d r.
\end{equation}
For the lattice version of the model, $a$~goes as $L^{-1}$, where~$L$ is
the (linear) system size, and hence the number of patches should scale as
$L^{d/(d+1)}$, i.e.,~as $L^{1/2}$ in one dimension or $L^{2/3}$ in two.
Numerical experiments on a one-dimensional system confirm this, giving a
measured exponent of $0.47\pm0.03$.

Now we wish to calculate the distribution $\rho(s)$ of fire sizes that
arises if we make this choice of patch sizes.  We have
\begin{equation}
\rho(s) = p(\vr) {\d^d r\over\d s}
        = p(\vr) {\d^d r\over\d p} {\d p\over\d s}
        = - ag{d+1\over d} p(\vr) {\d^d r\over\d p} s^{-(2+1/d)},
\label{pell}
\end{equation}
where $\d^d r$ here represents the volume of the space between the contours
$p$ and $p+\d p$ on the $p(\vr)$ surface.  As we will show, the term
$p(\vr)\,\d^d r/\d p$ is constant or contributes logarithmic corrections
for a wide selection of possible distributions $p(\vr)$, while the
principal power-law behavior in the event size distribution comes from the
factor~$s^{-(2+1/d)}$\footnote{As noted in Ref.~\onlinecite{DC00}, the
  exponent of this power law \emph{decreases} with increasing
  dimension~$d$, which is the opposite of the behavior seen in other models
  such as the self-organized critical forest fire model of B.~Drossel and
  F.~Schwabl, {\it Phys.\ Rev.\ Lett.}\ {\bf69}, 1629--1632 (1992).}.

An alternative method for deriving Eq.~\eref{pell} is to maximize the
simple yield functional $Y=1-\int p(\vr) s(\vr)\>\d^d r$, subject to a
constraint that fixes the volume $F$ occupied by the firebreaks
(Eq.~\eref{perimeter}).  This method, which is similar to the approach
taken in Ref.~\onlinecite{CD99}, is equivalent to the method above, via a
Lagrange transform, provided $F$ is chosen so as to make the corresponding
Lagrange multiplier equal to~$agd$.

Consider then the case (which covers all the examples in
Refs.~\onlinecite{CD99} and~\onlinecite{CD00}) of a distribution of sparks
with a single maximum at the origin, so that the volume $\d^d r$ takes the
form of an annulus enclosing the origin.  If we denote by $\Omega_d$ a
$d$-dimensional solid angle centered on the origin, then a volume element
of the annulus is $r^{d-1}\d r\,\d\Omega_d$.  In terms of the thickness $\d
p$ of the annulus, we can write $\d r = (r\,\d p)/(\vr\cdot\grad p)$, and
integrating our volume element over the contour $p=\mbox{constant}$ we find
\begin{equation}
{\d^d r\over\d p} = \oint_p {r^d\d\Omega_d\over\vr\cdot\grad p}.
\label{deriv1}
\end{equation}

For example Carlson and Doyle~\cite{CD99} studied the case of a spark
distribution in two dimensions having the form of the product of two
Gaussians with different widths:
\begin{equation}
p(\vr) = N
\exp\biggl(-\biggl[{x^2\over2\sigma_x^2}+{y^2\over2\sigma_y^2}\biggr]\biggr),
\label{gauss2d}
\end{equation}
where $N$ is a normalization constant.  For this distribution the
denominator of the integrand in Eq.~\eref{deriv1} is
\begin{eqnarray}
\vr\cdot\grad p &=&
  -N\biggl[{x^2\over\sigma_x^2}+{y^2\over\sigma_y^2}\biggr]
  \exp\biggl(-\biggl[{x^2\over2\sigma_x^2} + 
    {y^2\over2\sigma_y^2}\biggr]\biggr)\nonumber\\
  &=& 2p\log{p\over N},
\end{eqnarray}
which is constant over our contour of constant~$p$.  The element of solid
angle in two dimensions is simply the element of polar angle $\d\theta$,
and hence Eq.~\eref{deriv1} simplifies in this case to
\begin{equation}
{\d^2r\over\d p} = {1\over 2p\log(p/N)} \oint_p r^2\d\theta
                 = {A(p)\over p\log(p/N)},
\label{deriv2}
\end{equation}
where $A(p)$ is the area enclosed by the contour.  This contour is a line
of constant $(x/\sigma_x)^2+(y/\sigma_y)^2$, i.e.,~an ellipse, which has
major and minor axes $a = \sqrt{2\sigma_x^2\log(N/p)}$ and $b =
\sqrt{2\smash[b]{\sigma_y}^2\log(N/p)}$.  Thus the area enclosed by the
contour is $A(p) = \pi ab = 2\pi\sigma_x\sigma_y\log(N/p)$.  Combining
Eqs.~\eref{pell} and~\eref{deriv2} we then find that the distribution of
event sizes is
\begin{equation}
\rho(s) = 3\pi\sigma_x\sigma_y ag\,s^{-5/2}.
\end{equation}
Thus, for the Gaussian case in two dimensions the model generates a perfect
power-law with slope~$-\frac52$.

\begin{figure}
\resizebox{\figurewidth}{!}{\includegraphics{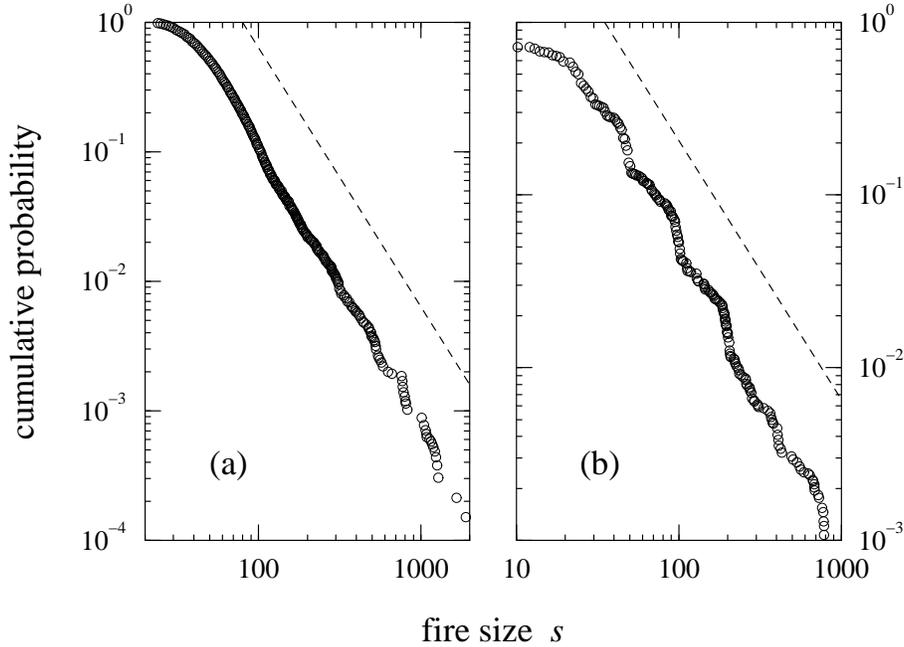}}
\caption{Cumulative distribution of fire sizes for simulations of the
  forest fire model in one and two dimensions, plotted on logarithmic
  scales.  (a)~One dimension with size~$10\,000$ and an exponential spark
  distribution.  (b)~Two dimensions with size $128\times128$ and a Gaussian
  spark distribution.  The results have been averaged over a number of
  different values of the parameters of the spark distributions to improve
  the statistics.  The dotted lines show the expected slopes of~$-2$
  and~$-\frac32$.}
\label{onetwo}
\end{figure}

This argument is easily generalized to other spark distributions and other
dimensions.  We find that the HOT forest fire model generates a perfect
power law with slope~$-(2+1/d)$ for all dimensions when a spark
distribution of the form $p(\vr) = N\exp(-\sum_{i=1}^d [x_i/\sigma_i]^d)$
is used.  As a test of this prediction, we show in Fig.~\ref{onetwo}
numerical results from direct simulations of the forest fire model in one
and two dimensions with distributions of this type~\footnote{The
  simulations were carried out using the greedy algorithm of
  Ref.~\onlinecite{CD99}, in which trees are placed on the lattice one by
  one, each in the position that maximizes the yield at that step.  This is
  not a true global optimization, but appears to give results close to the
  global optimum and in good agreement with theory.}.  For better
visualization and analysis the distributions pictured are cumulative, so
the expected slope is~$-(1+1/d)$, rather than~$-(2+1/d)$.  As the figure
shows, the slopes of the observed distributions are in good agreement with
this prediction.

We note in passing that the slope of~$-(1+1/d)$ for the cumulative
distribution of fire sizes seen in both our exact solution and our
numerical results is different from the slope of approximately~$-1$ found
numerically by Carlson and Doyle~\cite{CD00} in two dimensions.  The source
of this discrepancy is unclear, although it may be that the simulations of
Ref.~\onlinecite{CD00} provided too few data points to make an accurate
evaluation of the exponent possible.  We note also that the
value~$-\frac32$ for the two-dimensional case is quite different from the
slope of~$-\frac12$ measured for the cumulative size distribution of real
forest fires~\cite{DC00,MMT98}.

Other functions with exponential tails also generate power laws, but give
logarithmic corrections as well.  For instance, if $p(\vr) =
N\exp(-\sum_{i=1}^d [x_i/\sigma_i]^\gamma)$ with $\gamma\ne d$ then
Eq.~\eref{deriv2} still applies, but now $A(p)\sim[\log(N/p)]^{d/\gamma}$
and hence
\begin{equation}
p(\vr) {\d^d r\over\d p} \sim [\log(N/p)]^{d/\gamma-1}.
\end{equation}
Thus the distribution of event sizes fundamentally still follows a power
law with slope~$-(2+1/d)$, but there is a logarithmic correction.  Similar
logarithmic corrections are noted in Ref.~\onlinecite{CD99}.

Spark distributions with power-law tails also (unsurprisingly) give
power-law event size distributions, but in this case the exponent of the
distribution is non-universal, varying with the exponent of the spark
distribution.  For example, if $p(\vr)$ takes the generalized Lorentzian
form
\begin{equation}
p(\vr) = {N\over\sum_i (x_i/\sigma_i)^\nu+\Gamma^\nu},
\end{equation}
then we find
\begin{equation}
{\d^d r\over\d p} \sim {1\over p(\vr)}\,
                        \left({N\over p}-\Gamma^\nu\right)^{d/\nu-1},
\end{equation}
which goes asymptotically as $p^{-d/\nu}$ in the power-law tail where
$p(\vr)$ becomes small.  Thus the tail of the distribution of event sizes
goes as $\rho(s)\sim s^{-(2+1/d-d/\nu)}$.

We now turn to the COLD variant of the forest fire model, which
incorporates risk aversion.  In constructing this model we are guided by
theories of risk aversion in economics, where the subjective benefit of
outcomes is typically a nonlinear function of the loss~$s$, which is
captured by a utility function~$u(s)$~\cite{Savage72}.  Sensible utility
functions are decreasing with increasing loss: $u'<0$~\footnote{In
  traditional economics utility functions are usually defined to be
  increasing functions of their arguments; this could arranged in the
  present case, if necessary, by substituting $s\to-s$.}.  Risk aversion
also implies that $u''<0$, so that the negative utility of bad outcomes is
weighted more strongly than the positive utility of good outcomes.  One
standard family of utility curves that achieves this is the one-parameter
family
\begin{equation}
u(s) = {(1-s)^\alpha\over\alpha}.
\label{utility}
\end{equation}
Note that, since we will be concerned only with maximizing utility,
$u(s)$~is arbitrary to within both additive and multiplicative constants.
For $\alpha=1$, Eq.~\eref{utility} gives $u=1-s$ and maximizing utility is
precisely equivalent to minimizing loss.  For $\alpha<1$, we have
risk-averse utility functions and for $\alpha<0$ we are infinitely averse
to losing our entire investment.

Our goal now is to maximize the average utility functional
\begin{equation}
U = \int p(\vr) u(s(\vr)) \>\d^d r,
\end{equation}
subject to the constraint of fixed~$F$, Eq.~\eref{perimeter} (or
equivalently maximize a combined utility functional similar to
Eq.~\eref{yield}).  Carrying out the functional derivatives and using the
utility function of Eq.~\eref{utility}, we find that the optimum~$U$
corresponds to
\begin{equation}
p(\vr) {s(\vr)^{(d+1)/d}[1-s(\vr)]^{\alpha-1}} = \lambda,
\label{lagrange}
\end{equation}
where $\lambda$ is a Lagrange multiplier whose value can be calculated from
Eq.~\eref{perimeter}.  The distribution of event sizes is given by
Eq.~\eref{pell} as before, and using Eq.~\eref{lagrange} we find that the
derivative $\d p/\d s$, which gives the principal variation in~$\rho(s)$,
is
\begin{equation}
{\d p\over\d s} = \lambda\,{(\alpha+1/d)s-(1+1/d)\over
                   (1-s)^\alpha s^{2+1/d}}.
\label{dpds}
\end{equation}
For $\alpha=1$ our utility maximization is equivalent to simple yield
maximization (HOT), so it is not surprising to observe that when we set
$\alpha=1$ in the above expression we recover our previous $s^{-(2+1/d)}$
power-law.  For $\alpha<1$, we have risk-averse utility functions (COLD),
which give rise to event distributions following the $s^{-(2+1/d)}$ form
for small event sizes, but having lower probability of large event sizes.
When $\alpha<0$, event probability tends to zero as $s\to1$, as we would
expect.

\begin{figure}
\resizebox{\figurewidth}{!}{\includegraphics{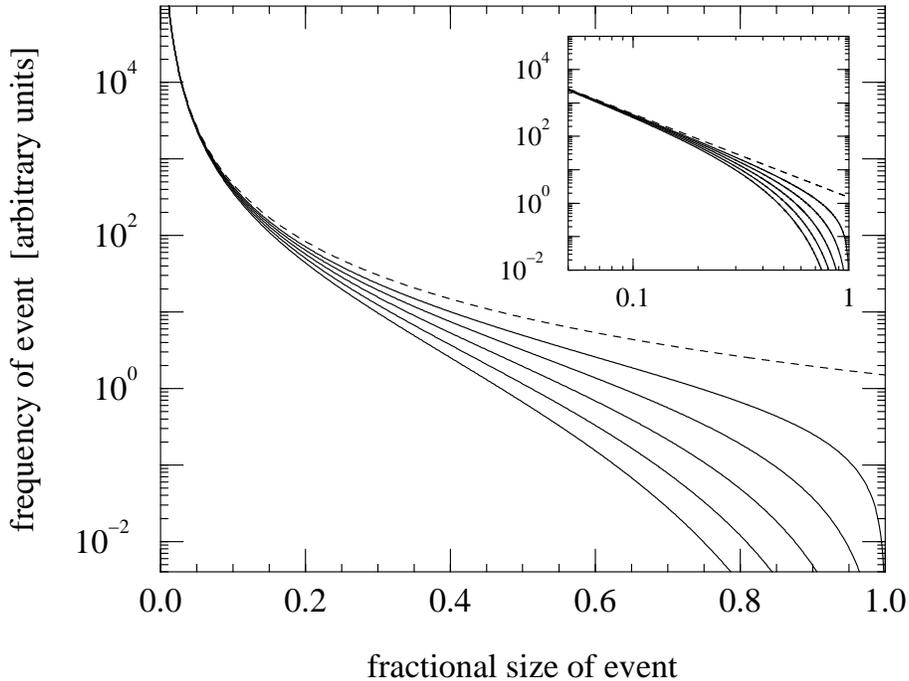}}
\caption{Event size distributions for HOT and COLD regimes in two
  dimensions.  The dotted line is the distribution for the HOT regime
  ($\alpha=1$) and the solid lines are for the COLD regime with (top to
  bottom) $\alpha=-1\ldots-5$.  The distributions are not normalized---they
  cannot be since they diverge at the origin.  In practice
  Eq.~\eref{perimeter} provides a lower cutoff on $s$ and makes the
  distribution normalizable.  Inset: the same data on log-log scales.}
\label{cold}
\end{figure}

In Fig.~\ref{cold} we compare the distribution of event sizes in HOT and
COLD regimes for a variety of values of the risk-aversion
parameter~$\alpha$.  The figure shows that the COLD distribution approaches
the HOT one as $\alpha$ approaches~1.  For $\alpha$ large and negative the
HOT power law is followed for only a small portion of the range of event
sizes---about 20\% in the case of $\alpha=-5$.

It is worth noting that while risk aversion truncates the power-law
behavior in the event size distribution, the distribution of the
\emph{utilities} of events still follows a power law: we find that the tail
of the distribution of utilities goes as $\rho(u)\sim u^{-\beta}$ with
$\beta=(2\alpha-1)/\alpha$.  Note that this exponent is independent of the
system dimension~$d$~\footnote{A similar result can also be derived within
  the framework of the ``PLR'' model of Ref.~\onlinecite{DC00}, which is a
  general model of engineered systems subject to stresses that occur with
  specified probability.  With this model a risk-averse utility function of
  the form~\eref{utility} again produces a sharp cutoff in the event
  distribution, but gives a distribution of the utilities of events that
  has a dimension-independent power-law tail.}.

While the introduction of the utility function has reduced the risk of
large losses, the optimal utility solution does not normally coincide with
the optimal yield solution, and hence we pay a cost for risk aversion in
terms of yield.  For the lattice forest fire model however we find that the
cost paid is small~\footnote{For other models or parameter regimes, such as
  the continuum version of the forest fire model with large~$a$ or the
  models of Ref.~\onlinecite{DC00}, the cost paid could certainly be
  larger.}.  For example, in a $128\times128$ two-dimensional system with a
Gaussian spark distribution, we find numerically that the mean yield at the
$\alpha=1$ optimum (HOT) is $0.904$, dropping to $0.900$ for $\alpha=-3$
and $0.888$ for $\alpha=-5$.  It appears therefore that the introduction of
risk aversion garners substantial benefits in terms of the reduction of
large losses---and the complete elimination of 100\% losses---while at the
same time costing us only a few percent at most in terms of average system
yield.

We conjecture that the suppression of power law tails in the COLD event
size distribution will also make the system more robust against the other
problems mentioned in the introduction, namely errors in the design and
changes in the spark distribution.  The truncation of the power law means
that the largest patches in the COLD solution are considerably smaller than
those in the HOT solution.  Thus, if a design flaw, such as a gap in one of
the firebreaks, causes two patches to merge, the resulting combined patch
is smaller too.  Similarly, if the spark distribution is changed, the size
of the resulting fires is smaller, and hence the effect on the average
yield is not as catastrophic as in the HOT case.  (A similar conjecture is
also made in Ref.~\onlinecite{CD99}.)

In this paper we have examined in detail only the forest fire model, but
similar principles should apply to other problems as well.  We have shown
that in order to produce power-law event size distribution, the HOT model
requires the auxilliary assumption of risk-neutrality.  If humans are
risk-averse they will tend to prefer COLD designs, although this does not
necessarily mean that HOT designs never occur.  It might be, for instance,
that blind evolutionary processes of the type found in natural systems
would simply optimize yield, without risk aversion.  On the other hand,
COLD designs are more robust to rare events than HOT designs, and therefore
might be selected for on long time-scales.  Of course, in the real world,
imperfect designs that fail to optimize either yield or utility are always
a possibility too.

{\bigbreak\small
The authors thank Jean Carlson and John Doyle for many useful comments and
suggestions.  This work was supported in part by the National Science
Foundation under grant number DMS--0109086 (MEJN and MG) and by McKinsey
Corporation, Credit Suisse First Boston, Bob Maxfield, and Bill Miller
(JDF).}

\end{document}